\documentclass[%
preprint,
 amsmath,amssymb,
 aps,
prb,
]{revtex4-2}
 
\usepackage{graphicx}
\usepackage{dcolumn}
\usepackage{bm}
\usepackage[thinc]{esdiff}
\usepackage{color}
\usepackage{mathrsfs}
\usepackage[normalem]{ulem}
\usepackage[dvipsnames]{xcolor}

\newcommand{\review}[1]{\textcolor{black}{#1}}

\begin{document}

The submitted manuscript has been created by UChicago Argonne, LLC, Operator of Argonne National Laboratory(“Argonne”). Argonne, a U.S. Department of Energy Office of Science laboratory, is operated under Contract No. DE-AC02-06CH11357. The U.S. Government retains for itself, and others acting on its behalf, a paid-up nonexclusive, irrevocable worldwide license in said article to reproduce, prepare derivative works, distribute copies to the public, and perform publicly and display publicly, by or on behalf of the Government.
\newpage

\raggedbottom    

\title{Extending the Takagi-Taupin equations for x-ray nanobeam Bragg coherent diffraction}

\author{T. Zhou$^1$}\email{tzhou@anl.gov}
\author{M.J. Cherukara$^2$}
\author{S. Kandel$^2$}
\author{M. Allain$^3$}
\author{N. Hua$^4$}
\author{O. Shpyrko$^4$}
\author{Y. Takamura$^5$}
\author{Z. Cai$^2$}
\author{S.O. Hruszkewycz$^6$}
\author{M.V. Holt$^1$}\email{mvholt@anl.gov}
\affiliation{
$^1$Center for Nanoscale Materials, Argonne National Laboratory, Argonne, Illinois 60439, USA}
\affiliation{
$^2$Advanced Photon Source, Argonne National Laboratory, Argonne, Illinois 60439, USA}
\affiliation{
$^3$Aix Marseille Univ, CNRS, Centrale Marseille, Institut Fresnel, Marseille, France}
\affiliation{
$^4$University of California San Diego, California 92093, USA}
\affiliation{
$^5$Department of Materials Science and Engineering, University of California, Davis, California 95616, USA}
\affiliation{
$^6$Materials Science Division, Argonne National Laboratory, Argonne, Illinois 60439, USA}
\date{\today}
  
\begin{abstract}
We present a new approach for simulating x-ray nanobeam Bragg coherent diffraction \review{patterns} based on the Takagi-Taupin equations. Compared to conventional methods, the current approach can be universally applied to any weakly strained system including semi-infinite crystals that diffract dynamically. It addresses issues such as \review{the curved wavefront and re-divergence} of the focused incident beam. We show excellent agreement against experimental data on a strained $\rm{La}_{0.7}\rm{Sr}_{0.3}\rm{MnO}_3$ thin film on $\rm{SrTiO}_3$ substrate, and a path to extracting physical information using automatic differentiation.
\end{abstract}


\maketitle

\section{Introduction}
The precise control of strain and crystalline quality is central to engineering desired properties across a wide array of functional materials. \review{Various} research goals, \review{ranging from} optimizing the electronic structure in hybrid perovskites \cite{Olthof2017Substrate-dependentPerovskites}, tuning the magnetotransport properties in magnetic oxides \cite{Yang2010StrainFilms}, \review{to} driving spin transitions in solid-state quantum systems \cite{Whiteley2019SpinphononAcoustics} all share the same material architecture that interfaces epitaxially with a high-quality single crystal substrate. Coherent diffraction imaging methods, despite recent advances in instrumentation \cite{Winarski2012AMicroscopy, Yan2019HardNm, Leake2019TheStructure} and methodology \cite{Schulli2018, Hruszkewycz2017High-resolutionPtychography, Miao2003PhaseMethod}, often shy away from this type of material system, due in large part to the inability to account for the substrates' contribution to the measured diffraction intensity. More specifically, conventional iterative phase retrieval methods rely on Fourier transform to describe the thin film diffraction process, which fails when the measured intensity is influenced by the strong substrate Bragg peak in close proximity.

Preliminary attempts to incorporate dynamical diffraction from semi-infinite crystals into nano-beam diffraction were successful\cite{Pateras2018DynamicalDiffraction}, but the Darwin's formalism based recursive calculation was limited to the case of planar thin films and to perfect substrates. The Takagi-Taupin equations (TTE)\cite{Takagi1962,Taupin1964} can be a good alternative. As a more general treatment of dynamical diffraction, the TTE is based on solving Maxwell's equations in a medium with a three-dimensional periodic distribution of dielectric susceptibility \cite{Ewald1916ZurKristalloptik, Laue1931DieForm, Bao2022}. While analytical solutions exist except for a few ideal cases \cite{Litzman1974TheCrystal, Katagawa1974TheEquations}, numerical calculations to the TTE can be performed on crystals of arbitrary shape \cite{Yan2014x-rayModeling}, having been extensively applied to calculate the reflectivity of bent single crystals \cite{Uschmann1993x-rayGeometry, Mocella2003AMonochromator, SanchezdelRio1997}. More recently, TTE has been applied to simulate Bragg Coherent Diffraction Imaging (BCDI) with an unfocused parallel beam \cite{Shabalin2017DynamicalCrystals, Hu2018DynamicImaging, Gao2022}. The studies showed dynamical diffraction artefacts in the far field for particles of size exceeding the extinction length, but offered no solution that can be embedded in the phase retrieval process.

In this work, we report the development of a TTE-based numerical framework for x-ray nanobeam Bragg coherent diffraction that tackles key issues related to the use of a convergent coherent beam and to compatibility with phase retrieval approaches. The incident \review{angular} spread of a tightly focused beam, combined with the large acceptance of two dimensional detectors, is known to satisfy multiple diffraction conditions at once \cite{Holt2013NanoscaleStudies}. The diffraction scenario is thus more complex than with a parallel beam as in BCDI or reflectivity experiments. Additionally, the propagation of the curved wavefront, more specifically the re-divergence of the focused beam also needs to be taken into account for sample thickness beyond the depth of focus.

We examine the validity of our approach by a quantitative comparison between the model-predicted far field diffraction patterns and those obtained in experiments. The current approach differs from conventional Fourier transform based methods mainly by its direct expression of the lattice strain and its ability to account for the dynamically diffracting substrate. Its explicit description of the scattering condition has proven useful in uniting datasets acquired at different sample and detector angles. More importantly, the close resemblance of the numerical framework to an artificial neural network (NN) indicates that any optimizers for NN training can be applied to perform the inverse process (\textit{e.g.,} from diffraction pattern to strain in the sample). We demonstrate this \review{capability} by simultaneously retrieving the lattice strain and film thickness from a single x-ray nanobeam Bragg coherent diffraction pattern using gradient descent optimization with automatic differentiation. \review{The inverse process demonstrated in this work is an important first step to achieve phase retrieval in scanning x-ray nanobeam coherent diffraction, which is an emerging technique colloquially known as Bragg Ptychography\cite{Chamard2015, Hruszkewycz2017High-resolutionPtychography}.}\\\indent 

\section{Experimental Method and Simulation Framework}

\subsection{Bragg coherent diffraction of a thin film}
\begin{figure}[bt]
\centering
\includegraphics[width=1\textwidth]{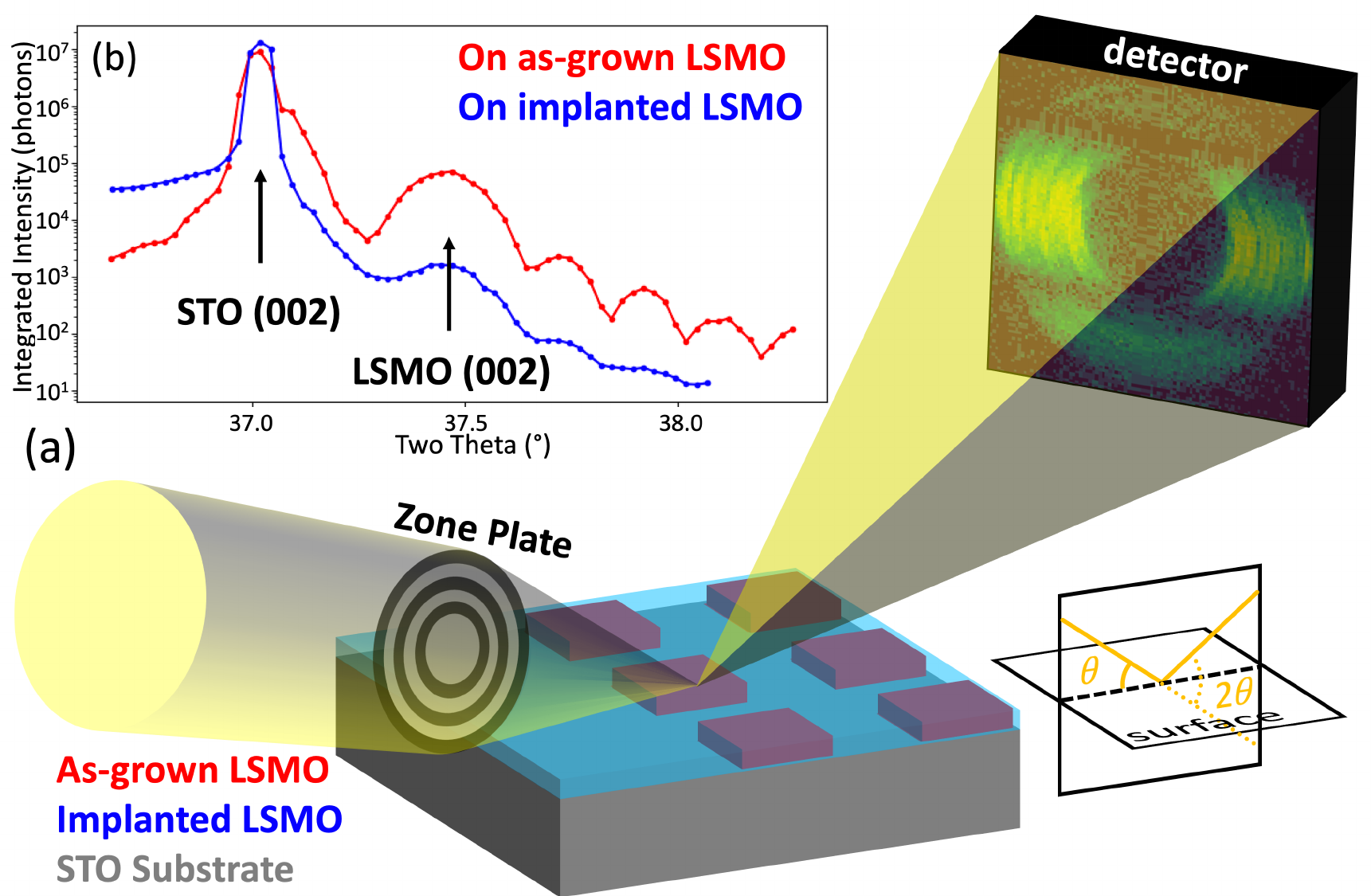}
\caption{\label{fig:exp_sketch} (a) Schematic of the experimental setup. Zone plate optics were used for the focusing of the incident beam. A cropped area (100$\times$100 pixels) of the detector image is shown here for clarity. (b) Specular rod profile extracted respectively from the $\theta$-2$\theta$ scan on as-grown and implanted LSMO.}
\end{figure}

The sample studied in this work was an epitaxially strained $\rm{La}_{0.7}\rm{Sr}_{0.3}\rm{MnO}_3$ (LSMO) thin film ($\approx$~40 nm thickness) grown on a (001)-oriented $\rm{SrTiO}_3$ (STO) substrate, and selectively patterned by $\rm{Ar}^+$ implantation \cite{Takamura2006TuningIslands}. (Fig.~\ref{fig:exp_sketch}) Through lithographic masking, arrays of square pads of intact crystalline LSMO film, 2$\times$2 $\mu m^2$ in size separated by 2 $\mu m$, were fabricated. Regions of LSMO outside of the masked pad areas were partially amorphized by the ion implantation.

X-ray nano-diffraction measurements of this sample were performed on the 26-ID-C beamline at the Advanced Photon Source (Fig.~\ref{fig:exp_sketch}a). The focused beam size was 60 nm at 10 keV, as confirmed by transmission ptychographic reconstructions on a test pattern. Two sets of data were acquired by separately positioning the focused beam in the center of a masked pad and in the implanted area. For each dataset, a radial $\theta$-2$\theta$ scan \review{of 65 points} was performed about the \review{specular} 002 LSMO and STO Bragg reflections. \review{The total scattering angle $2\theta$ was around 37$^\circ$.} The radial scans consisted of moving the sample angle in increments of $\Delta\theta=0.0125^\circ$, while shifting simultaneously the detector angle by 2$\Delta\theta$. \review{The scattering plane was horizontal, and the rotations of both the sample and the detector were around a vertical axis.} Each point of the radial scan \review{was} associated with a far-field diffraction pattern, measured on a Medipix detector (55 $\mu m$ pixel size, 515$\times$515 pixels) at a sample-to-detector distance of 0.9 m. 

The $\theta$-2$\theta$ \review{motion} scanned the momentum transfer along the specular rod 00$\ell$ in reciprocal space. Fig.~\ref{fig:exp_sketch}b shows the specular rod profile for the two data sets generated by integrating the intensity in a narrow region of interest (RoI) centered on the detector. The strong intensity oscillations on the red curve confirmed that the masked area consisted of an as-grown LSMO film of high crystal quality while the dampened intensity oscillation on the blue curve indicated that the LSMO in the implanted area was significantly damaged. 

\subsection{Extending the TTE framework for x-ray nanobeam Bragg coherent diffraction}

\review{To simulate x-ray nanobeam Bragg coherent diffraction patterns, we have adopted} the symmetric version of two beam TTE \cite{Shabalin2017DynamicalCrystals}:
\begin{equation}
\label{eq_tte}
\begin{split}
\diffp{E_{0}(\mathbf{r})}{{s_0}} = \frac{i\pi}{\lambda}[\chi_0 E_{0}(\mathbf{r})+P\chi_{\overline{h}}e^{-i\Delta\mathbf{h}\cdot\mathbf{r}+i\mathbf{h}\cdot\mathbf{u(r)}}E_{h}(\mathbf{r})]\\
\diffp{E_{h}(\mathbf{r})}{{s_h}} = \frac{i\pi}{\lambda}[\chi_0 E_{h}(\mathbf{r})+P\chi_he^{i\Delta\mathbf{h}\cdot\mathbf{r}-i\mathbf{h}\cdot\mathbf{u(r)}}E_{0}(\mathbf{r})].
\end{split}
\end{equation}

\review{Here, ${E_{0}(\mathbf{r})}$ and ${E_{h}(\mathbf{r})}$ are respectively the scalar pseudo amplitudes for the transmitted wave $\mathbf{E_0}$ and the diffracted wave $\mathbf{E_h}$. Their partial derivatives are taken with respect to $s_0$ and $s_h$, which are unit distances along the directions of their respective wave vectors $\mathbf{K_0}$, and $\mathbf{K_h}$.} $i$ is the imaginary unit. $\lambda$ is the wavelength of the x-ray beam. The Fourier components of the dielectric susceptibilities $\chi_0, \chi_h, \chi_{\overline{h}}$ \review{are complex quantities related to the structure factor at their corresponding reciprocal lattice vectors as denoted by the subscripts.} $P$ is the polarization factor. \review{For specular reflections in the horizontal scattering geometry such as those measured in this work, $P=\cos2 \theta$.} $\mathbf{u(r)}$ is the strain field. $\mathbf{h}$ is the \review{reciprocal lattice vector}. $\Delta\mathbf{h}$ is the \review{deviation} from the Bragg condition \review{at $\mathbf{h}$.} A \review{formula} of $|\Delta \mathbf{h}|$ for rocking scans can be found in \cite{Shabalin2017DynamicalCrystals}. For the radial $\theta$-2$\theta$ scans around a specular reflection as explored in this work, we \review{derive the following formula}
\begin{equation}
|\Delta \mathbf{h}| = \frac{4\pi}{\lambda}\Delta\theta\cos\theta.
\label{eq_radial}
\end{equation}

\begin{figure}[bt]
\centering
\includegraphics[width=1\textwidth]{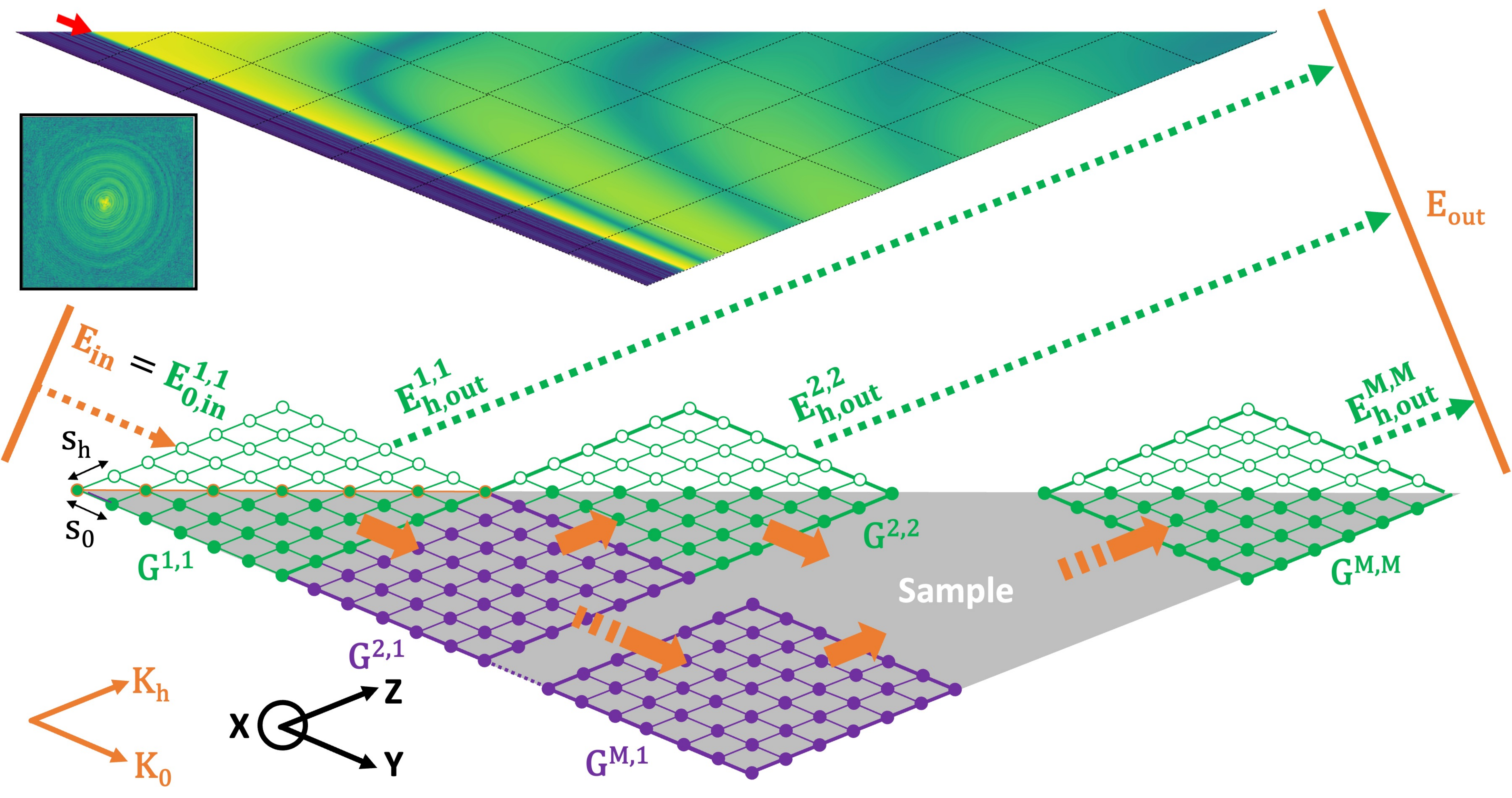}
\caption{\label{fig:sim_sketch} Bottom: Cross-sectional schematic of the \review{numerical} framework used in this work. The orange arrows indicate the direction of wave propagation between the adjacent \review{subsections}. Top: Cross-sectional view of \review{the amplitude of} $E_h$ in logarithmic scale, simulated using 36 parallelepiped-shaped subsections for a total substrate thickness of $8.76 ~\mu m$. The step size for numerical integration is $s_0 = s_h = 13.16 $ nm. The red arrow indicates the direction and point of incidence of the nano-focused beam. (inset) \review{Amplitude of} the initial condition $E_{in}$ in logarithmic scale.}
\end{figure}

The scheme for solving the TTE in this work is illustrated in Fig.~\ref{fig:sim_sketch}. For information on the numerical integration of the TTE using the half-step method, the reader is referred to the original work of Authier\cite{Authier1968EtudeDeforme} and Gronkowski\cite{Gronkowski1991}. In this section, we focus on the differences between this work and those previously reported for simulating BCDI images\cite{Shabalin2017DynamicalCrystals, Hu2018DynamicImaging, Gao2022}. \review{Specifically, the application of BCDI is limited to isolated objects with size smaller than the illumination, whereas there is in principle no size limit for experiments with a scanning x-ray nanobeam, laterally or in the depth dimension. The extensions described in this work mostly deal with problems arising from wave propagation in a large probed volume, namely the re-divergence and the curved wavefront of the focused incident beam.}

The \review{numerical integration} volume is divided into \review{parallelepiped-shaped} subsections denoted \review{$G^{m,n}$}. \review{The $\mathbf{Y}$ and $\mathbf{Z}$ axes of the parallelepipeds are chosen to be along respectively the direction of the incident wavevector $\mathbf{K_0}$ and the diffracted wavevector $\mathbf{K_h}$. The $\mathbf{X}$ axis of the parallelepipeds is perpendicular to the scattering plane spanned by $\mathbf{Y}$ and $\mathbf{Z}$. The superscripts $m$ and $n$ denote the number of subsections in which the exit wave would have propagated along respectively the direction of $\mathbf{K_0}$ and $\mathbf{K_h}$. For a given subsection $G^{m,n}$ as shown in Fig.~\ref{fig:sim_sketch}, its top-left and bottom-right edges are respectively the incident and exit surface for $\mathbf{E_0}$, while its bottom-left and top-right edges are respectively the incident and exit surface for $\mathbf{E_h}$. The wave at the incident and exit surfaces are denoted with subscripts $in$ and $out$.} The first parallelepiped $G^{1,1}$ contains the illuminated surface area. \review{To account for the curve wavefront of the focused incident beam, the initial condition is set as
\begin{equation}
\label{eq_init1}
E_{0,in}^{1,1} = E_{in}.
\end{equation}
$E_{0,in}^{1,1}$ is the transmitted wavefield $E_0$ at the incident surface of $G^{1,1}$. 
The incident wavefield $E_{in}$ is obtained through probe reconstruction with transmission ptychography on a resolution target, under otherwise the same illumination condition. An example of $E_{in}$ is shown in the inset of Fig.~\ref{fig:sim_sketch}. The remaining initial conditions are
\begin{equation}
\label{eq_init2}
E_{0,in}^{m,m} = 0,\; m = 2, 3,...
\end{equation}
\begin{equation}
\label{eq_init3}
E_{h,in}^{m,1} = 0,\; m = 1, 2,...
\end{equation}
$G^{m,m}$ correspond to the subsections containing part of the sample surface, and are marked as green parallelograms in the schematic shown in Fig.~\ref{fig:sim_sketch}. Eq.~(\ref{eq_init2}) indicates that, except for $G^{1,1}$, there is no incident beam on the sample surface. $G^{m,1}$ are the subsections in the path of the incident beam, and are marked as purple parallelograms in the schematic shown in Fig.~\ref{fig:sim_sketch}. Eq.~(\ref{eq_init3}) indicates the absence of diffracted wave entering the left-most edge of the integration volume.}

The procedures for wave propagation inside each subsection follows the same principle as described in the demonstrations for BCDI\cite{Shabalin2017DynamicalCrystals, Hu2018DynamicImaging, Gao2022}. Numerical integration of Eq.~(\ref{eq_tte}) is performed sequentially along the beam propagation directions $\mathbf{Y}$ and $\mathbf{Z}$. The calculation along $\mathbf{X}$ is concurrent. \review{Once the calculation on a subsection $G^{m,n}$ is complete, the wavefields $E_{0,out}^{m,n}$ and $E_{h,out}^{m,n}$ at its exit surfaces are collected to serve as the boundary conditions for subsequent calculations, with
\begin{equation}
\label{eq_bound}
\begin{split}
E_{0,in}^{m+1,n} = E_{0,out}^{m,n}\\
E_{h,in}^{m,n+1} = E_{h,out}^{m,n}.
\end{split}
\end{equation}
}
Fig.~\ref{fig:sim_sketch} also shows a cross-section of $E_h$ in the scattering plane \review{spanned by $\mathbf{Y}$ and $\mathbf{Z}$}, calculated at \review{the Bragg condition of the substrate STO 002 reflection}. Dynamical effects are clearly visible as evidenced by the Pendellösung fringes. 

Breaking down the integration network into smaller subsections \review{has the advantage of reducing the computer memory needed for numerical integration on an exceedingly large sample volume. Figure~\ref{fig:order} in \textbf{Appendix A} depicts in details the order of computation for each sub-section $G^{m,n}$ in the case of $M=5$. The boundary conditions between neighbouring sub-sections are listed explicitly. These boundary conditions are what is limiting the calculations from being carried out simultaneously on all subsections. More importantly, the division into smaller subsections allows, to an extent, consideration of the re-divergence of the focused incident beam.} To achieve this, we draw inspiration from the multi-slicing technique commonly used in electron diffraction \cite{Cowley1995}. The wave exiting one \review{subsection} is Fresnel propagated before entering the next. We note that the distance of the Fresnel propagation ($\sim$1~$\mu$m, the \review{side length} of a \review{subsection} along the propagation direction) is much smaller compared to the depth of focus of most nano-focusing x-ray optics ($\sim$10 $\mu$m).

The final diffracted intensity $I$ is obtained by far-field propagating the exit wave ${E_{out}}$, which is itself obtained by concatenating all the diffracted wavefield $E_{h,out}^{m,m}$ exiting the sample surface. 
\begin{equation}
I  = |\mathscr{F}({E_{out}})|^2,
\label{eq_I}
\end{equation}
where $\mathscr{F}$ denotes the Fourier transform, and
\begin{equation}
{E_{out} = {E_{h,out}^{1,1}} {^\frown} {E_{h,out}^{2,2}} {^\frown} ... {^\frown} {E_{h,out}^{M,M}}}
\label{eq_Eout},
\end{equation}
where ${^\frown}$ denotes concatenation. 

\section{Results}
\subsection{\review{Nanobeam} coherent diffraction at the thin film Bragg peak}
\begin{figure}[bt]
\includegraphics[width=1\textwidth]{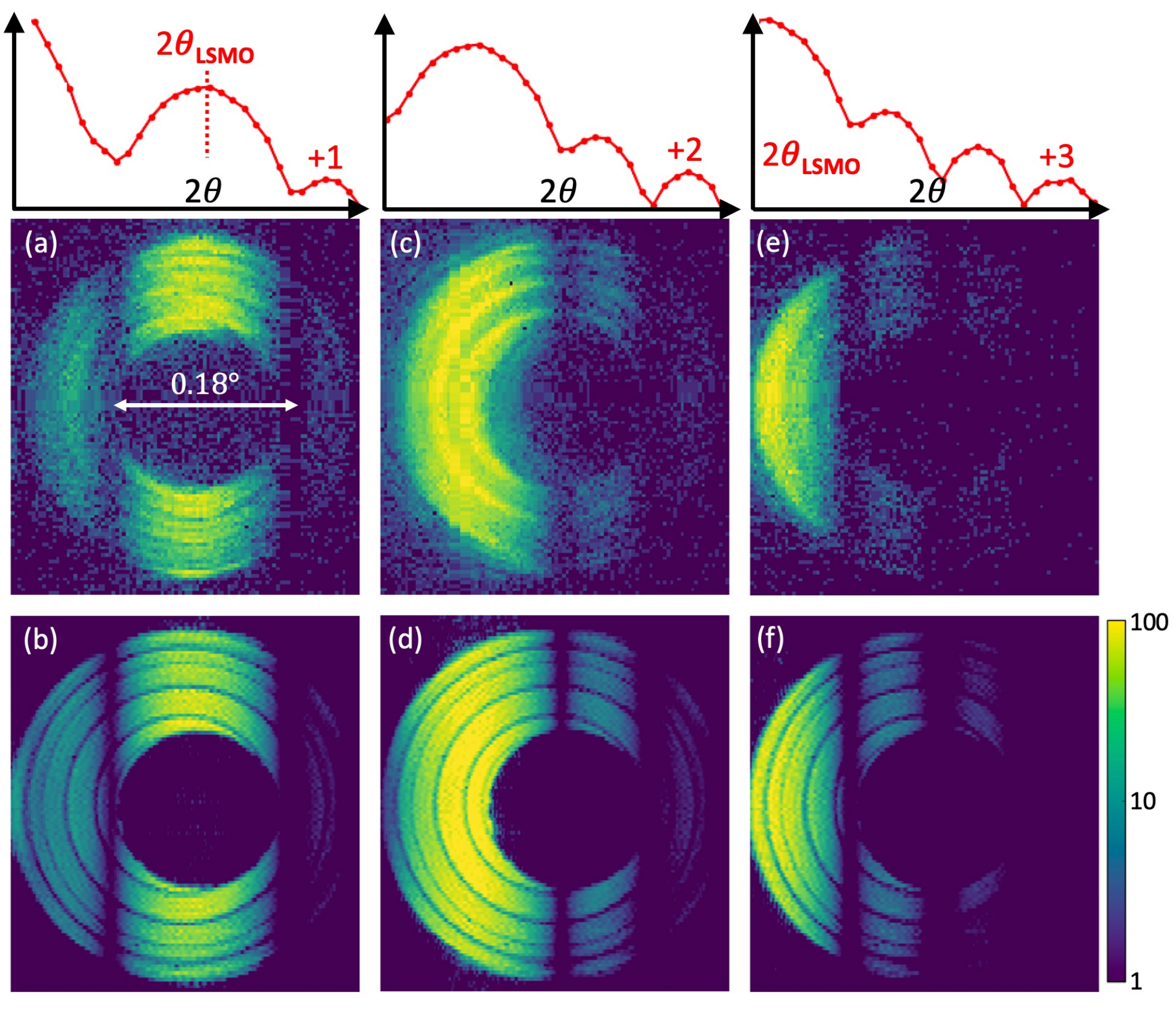}
\caption{\label{fig:film_peak} Comparison between experimental (exp) and simulated (sim) far-field diffraction patterns at various incident angles. The beam was focused on the center of a masked (as-grown) LSMO pad. (a) exp and (b) sim data at $\theta=\theta_{\rm{LSMO}}$. (c) exp and (d) sim data at $\theta=\theta_{\rm{LSMO}}+0.1^\circ$. (e) exp and (f) sim data at $\theta=\theta_{\rm{LSMO}}+0.2^\circ$. Their corresponding sections of the specular rod profile are shown in the insets above with the numbering -1, +1, +2 and +3 marking the position of the maxima of different orders of the thickness fringes. The intensity scale bar applies to all images.}
\end{figure}
We first show how the proposed TTE-based numerical framework can accurately model thin film coherent diffraction with a nano-focused beam. For a single nanodiffraction image acquired at a given incident angle, a significant area of the reciprocal space was simultaneously measured due to the large acceptance of the area detector and to the large convergence angle of the focused beam. This is illustrated in Fig.~\ref{fig:film_peak}a, which shows a diffraction pattern acquired when the focused beam was centered on the masked LSMO pad. The sample angle of this measurement was such that the LSMO 002 Bragg condition was satisfied ($\theta= \theta_{\rm{LSMO}}$). 
The section of the specular crystal truncation rod (CTR) profile corresponding to the $2\theta$ range of the detector image is plotted in the inset directly above Fig.~\ref{fig:film_peak}a. The relationship between the nanodiffraction pattern and the CTR is as follows: the minima of the CTR intensity oscillations appear as dark stripes on the nanodiffraction pattern, while the film Bragg peak corresponds to the bright band in the middle of the detector, which is occluded by the shadow of the central beam stop. Another important feature to note is that the left (smaller 2$\theta$) side of the diffraction pattern is brighter than the right side (larger 2$\theta$). This asymmetric intensity distribution is caused by the strong 002 STO substrate Bragg peak, which diffracts at a slightly smaller $\theta/2\theta$ angle ($\theta_{\rm{STO}}=\theta_{\rm{LSMO}}-0.23^\circ$).\\\indent
This experimental image, which encompasses simultaneously nanodiffraction from a thin film and contribution from the substrate, offers a unique opportunity to validate our forward model. Fig.~\ref{fig:film_peak}b shows the simulated diffraction pattern obtained by varying the thin film parameters until it fit best to the experimental data. This in turn allowed us to determine the thickness of the LSMO film to be $41\pm3$ nm and its out-of-plane lattice parameter to be $c_{\rm{LSMO}}^{\rm{obs}}=3.858$\AA. Comparing with the bulk lattice parameter of $c_{\rm{LSMO}}^{\rm{bulk}}=3.8935$\AA \cite{Vailionis2011MisfitModulations}, we found that the film was under an out-of-plane compressive strain of -0.91\%. This strain state is indeed expected due to the lattice mismatch of the epitaxial LSMO film with the STO substrate ($a=3.905$ \AA) and was previously observed with surface x-ray diffraction measurement \cite{Herger2008StructureMagnetoresistance, Vailionis2014SymmetryInterfaces}. Excellent predictions were also obtained against the experimental data at larger $\theta$/2$\theta$ angles (Fig.~\ref{fig:film_peak}c and \ref{fig:film_peak}e) measured during the radial scan. It is worth mentioning that \review{to produce} the results shown in Fig.~\ref{fig:film_peak}d and \ref{fig:film_peak}f, \review{it sufficed to change just the $\Delta\theta$ variable in Eq.~\ref{eq_radial}}. The same parameters determined previously for Fig.~\ref{fig:film_peak}b were used, without extra fitting or rescaling of the intensity level.\\\indent
\review{Our numerical framework} correctly replicated the asymmetric intensity distribution observed in Fig.~\ref{fig:film_peak}a, with the simple assumption of a perfectly crystalline (001) oriented STO substrate with thickness \review{larger than} the Bragg extinction \review{depth. Fig.~\ref{fig:coupling}a of \textbf{Appendix B} shows a comparable diffraction pattern from the thin film alone. A more symmetric intensity distribution was observed which was expected in the absence of the substrate Bragg peak. More importantly, our numerical framework correctly accounted for the coupling between the diffraction from the thin film and from the substrate. Fig.~\ref{fig:coupling}b shows a comparison between the intensity profile with and without such coupling. With the film-substrate coupling, the simulated intensity using the numerical framework presented in this work matched well with the experimental data. Without the film-substrate coupling, the intensity of the -1 order thickness fringe was under-estimated. Without the film-substrate coupling, we also observed a slight shift of the center of the LSMO Bragg peak to larger $2\theta$ values by 0.003$^\circ$, which could in turn be wrongly interpreted as a slightly larger compressive strain by 0.01\%.}\\\indent

\subsection{\review{Nanobeam} coherent diffraction at the substrate Bragg peak}
\begin{figure}[b]
\centering
\includegraphics[width=1\textwidth]{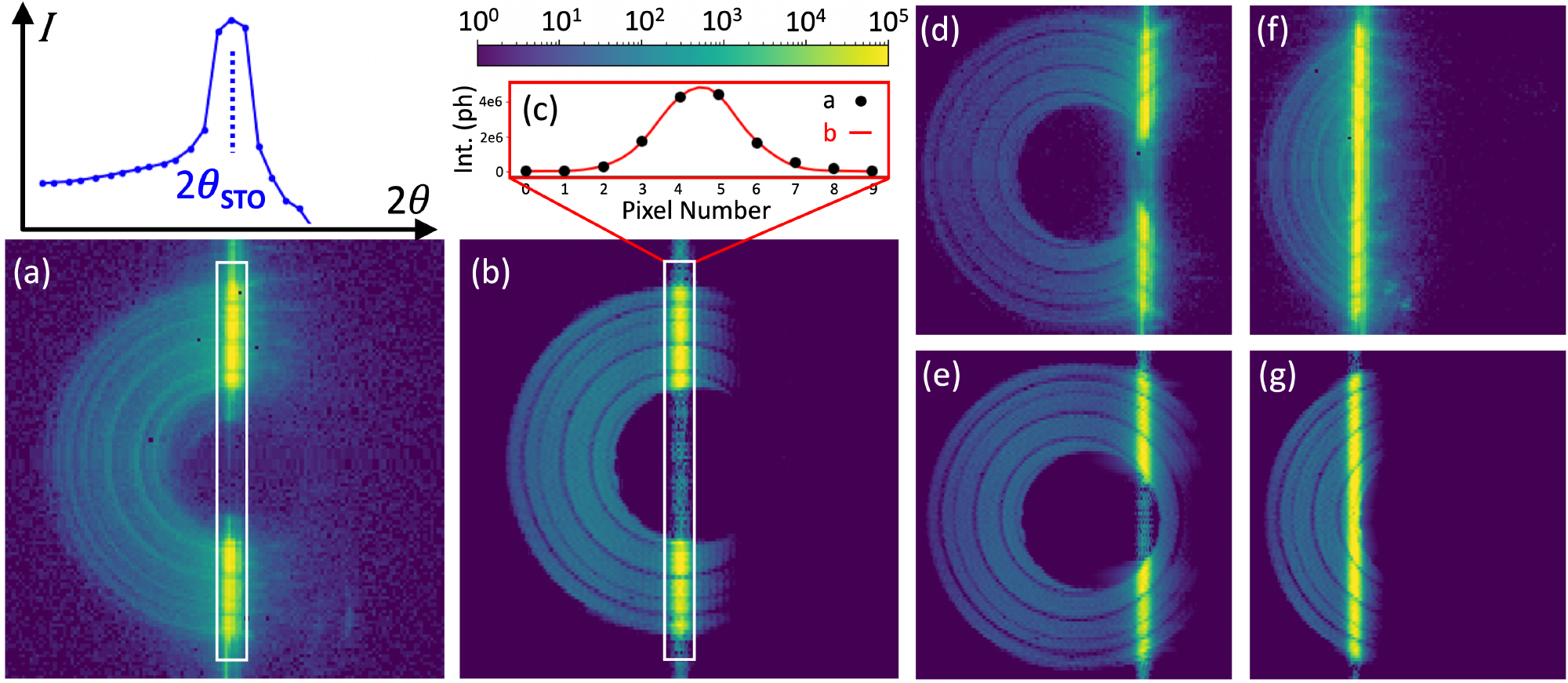}
\caption{\label{fig:sub_peak} Comparison between experimental (exp) and simulated (sim) far-field diffraction patterns at various incident angles. The beam was focused on implanted LSMO far away from the masked array. (a) exp and (b) sim data at $\theta=\theta_{\rm{STO}}$. (c) Width of the substrate Bragg peak in pixels. The data was extracted from the white box area in (a) and (b) then summed along the vertical axis. (d) exp and (e) sim data at $\theta=\theta_{\rm{STO}}-0.075^\circ$. (f) exp and (g) sim data at $\theta=\theta_{\rm{STO}}+0.075^\circ$. The intensity scale bar applies to all images.}
\end{figure}
Accurate model prediction was also achieved in close vicinity to the substrate Bragg peak. This is demonstrated for data measured from the ion-implanted region of the sample. Fig.~\ref{fig:sub_peak}a shows the experimental diffraction pattern taken at the diffraction condition of the STO 002 reflection ($\theta= \theta_{\rm{STO}}$), along with the corresponding section of the specular CTR profile.  The bright vertical streak in the middle of the image is the STO substrate 002 Bragg peak, occluded by the shadow of the central beam stop. The ``plateau'' of weak intensity on its left (smaller 2$\theta$) side originates from contribution from the irradiated LSMO film layer. In order to produce the best-fit image in Fig.~\ref{fig:sub_peak}b, the out-of-plane lattice parameter of the LSMO used in the simulation was adjusted to $c_{\rm{LSMO}}^{\rm{ion}}=3.940$\AA, which corresponds to an out-of-plane tensile strain of 1.20\%. Such lattice expansion has previously been observed in $\rm{Ar}^+$ implanted LSMO layer \cite{Takamura2006TuningIslands, Lee2020}.\\\indent
We note that dynamical effects alone cannot explain the observed \review{broadening} of the Bragg streak along the $2\theta$ direction. \review{As shown in Fig.~\ref{fig:sub_peak}c, the experimentally observed peak breadth of 0.008$^\circ$ FWHM is much larger than the Darwin width of a dynamically diffracting STO 002 peak, calculated at 0.002$^\circ$ FWHM.} The difference between the experimental and simulated peak breadth is attributed to the mosaicity in the substrate. \review{In \textbf{Appendix C}, we describe a strategy to properly account for the effect of substrate mosaicity in the simulation. Using this strategy and} the material parameters \review{determined in Fig.~\ref{fig:sub_peak}b}, very good agreement was also found between experiment (Fig.~\ref{fig:sub_peak}d and \ref{fig:sub_peak}f) and the model prediction (Fig.~\ref{fig:sub_peak}e and \ref{fig:sub_peak}g) at incrementally different $\theta/2\theta$ positions, without extra fitting or rescaling of the intensity level.\\\indent

\subsection{Extracting sample information from a single diffraction pattern}
\begin{figure}[bb]
\centering
\includegraphics[width=1\textwidth]{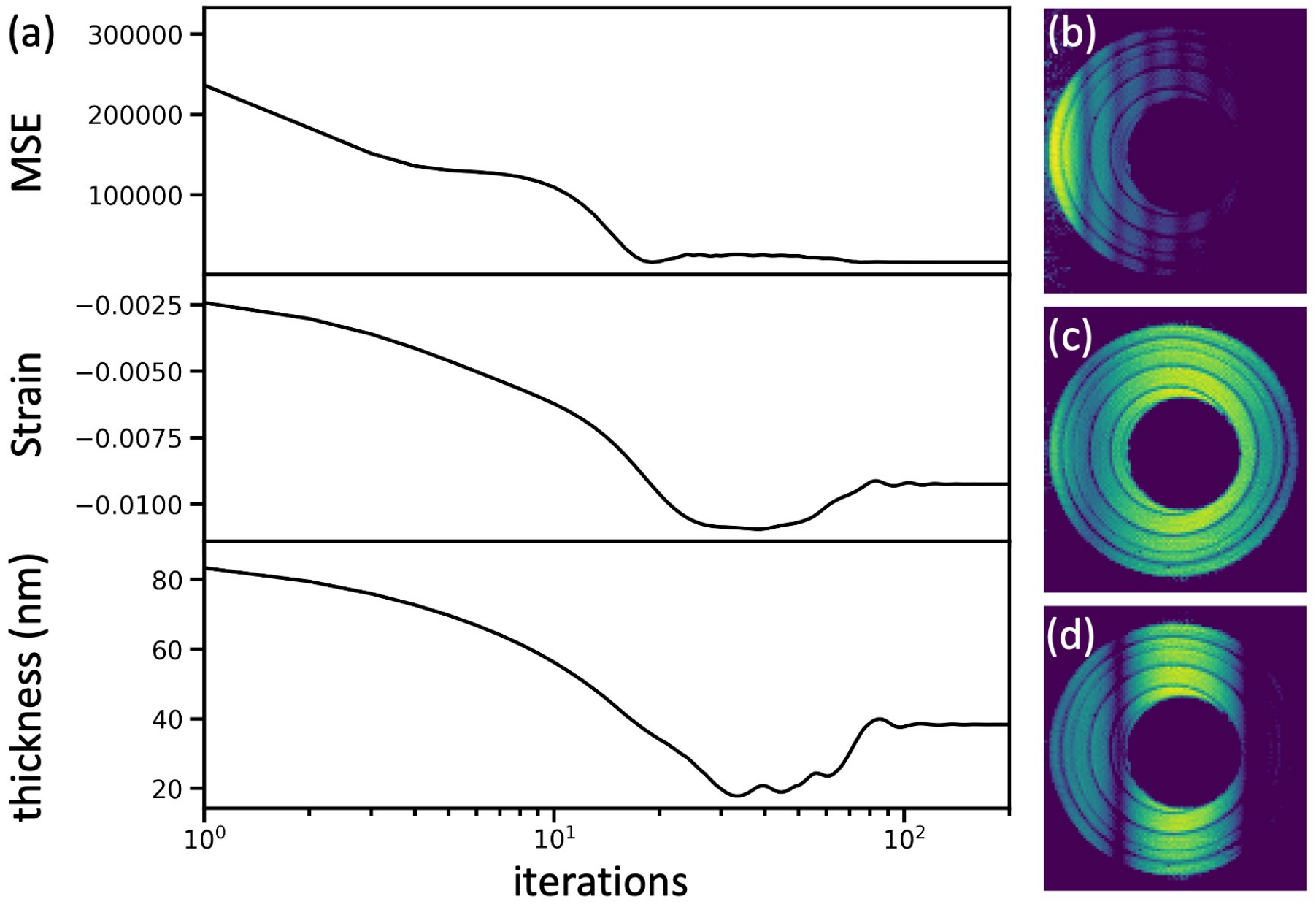}
\caption{\label{fig:AD} (a) Evolution of mean squared error, retrieved strain and film thickness. The corresponding diffraction pattern at iteration number (b) 0, (c) 50 and (d) 100. The intensity is shown in the same logscale as Fig.~\ref{fig:film_peak}b.}
\end{figure}
The good agreement obtained against \review{the} experimental data makes the proposed method a promising candidate for extracting physical information of the sample. Here we show a demonstration of retrieving the thickness and out-of-plane strain of the LSMO layer directly from individual diffraction patterns, assuming the film to be uniform within the illuminated area. Such a fitting approach can also be scaled up to the case of multiple beam positions across the sample to enable phase retrieval in scanning nanobeam Bragg coherent diffraction, also known as Bragg Ptychography\cite{Chamard2015, Hruszkewycz2017High-resolutionPtychography}. 
The principle behind the parameter retrieval from a single diffraction pattern is similar to what was previously demonstrated for Lorentz transmission electron microscopy\cite{Zhou2021}. A simulated diffraction pattern was first generated using a set of guessed parameters through the forward propagation model described in \textbf{section II.B}. The mean squared error (MSE) between the simulated and experimental diffraction pattern was then calculated. Subsequently, the gradient of the MSE with respect to the input parameters is computed using reverse mode automatic differentiation \cite{Kandel2019, Nashed2017DistributedPtychography}, and the parameters are then updated in steps proportional to the negative of the gradients. From there, the entire process is repeated until the improvement of the MSE falls below a predefined value. \\\indent
We demonstrated retrieval of both the strain and the thickness of the LSMO film from an individual diffraction pattern using the experimental data shown in Fig.~\ref{fig:film_peak}a. The initial guess for the two parameters were respectively -0.2\% and 90 nm, to show how the proposed scheme perform under initial conditions far away from the ground truth.
Fig.~\ref{fig:AD}a shows the evolution of the MSE, strain and film thickness. The correct values were obtained at about 100 iterations, after which both the MSE and the parameters ceased to change. Fig.~\ref{fig:AD}b shows the diffraction pattern simulated with the initial guessed parameters. Compared to the experimental (target) data, the spacing between the bright and dark stripes was too small due to an overestimated film thickness. The brightest band was found too much to the left (smaller $2\theta$) side of the image due to an underestimated compressive strain. After 50 iterations (Fig.~\ref{fig:AD}c), the retrieved strain was close to the actual value which explains the roughly correct position of the bright band, but the retrieved thickness overshot to a smaller value, resulting in the larger width of the thickness fringes. Fig.~\ref{fig:AD}d shows the diffraction pattern simulated using the parameters retrieved after 100 iterations, with good agreement with the experimental data shown in Fig.~\ref{fig:film_peak}a. The small discrepancy between the final optimized diffraction pattern and the simulated diffraction pattern shown previously in Fig.~\ref{fig:film_peak}b originates from the smaller substrate volume (1 $\mu$m thickness) being considered while performing the parameter retrieval. We use the ADAM optimizer\cite{Kingma2015} implemented in Google's Tensorflow package. The initial learning rate was set to 1.  Each iteration took 90 sec on an Intel Xeon E7 CPU.\\\indent

\section{Conclusion}
To summarize, the Takagi-Taupin equation was extended to simulate x-ray nanobeam Bragg coherent diffraction data. Excellent agreement was obtained against experimental data taken on epitaxially strained and on ion implanted LSMO thin film grown on STO substrate, and characteristics of the thin film were determined via gradient descent optimization enabled by automatic differentiation.\\\indent
A key feature of the current development is its ability to account for substrates' contribution to the measured diffraction intensity. We demonstrate this by correctly reproducing the asymmetric intensity distribution observed near the LSMO thin film Bragg peak in Fig.~\ref{fig:film_peak}. We note that one does not need to be in close vicinity of the substrate peak for its influence to be significant. The asymmetric intensity distribution was observed at more than $0.2^\circ$ away in $\theta$ from the substrate Bragg condition, which is over 25 times the width of the substrate peak (FWHM = $0.008^\circ$). We further note that due to the complex diffraction conditions excited simultaneously by the large convergent angle of the incident beam and by the large acceptance of the two-dimensional detector, the dynamically diffracting substrate peak was observable over a much larger angular range in nano-beam diffraction than in parallel beam diffraction (including BCDI) as shown in Fig.~\ref{fig:sub_peak}.\\\indent
Another important benefit of the current approach is its explicit description of the scattering condition $\Delta \mathbf{h}$ and the strain field $\mathbf{u(r)}$. The former enables the modeling of experimental data taken at different sample and detector angles with the change of one single parameter, which is convenient when multiple datasets on the same sample need to be together processed such as in the case of multi-angle Bragg Ptychography \cite{Hill2018}. The latter is interesting as it connects the measured intensity directly to strain, bypassing the cryptic phase information as seen in many phase retrieval methods. Additional steps were also implemented in the \review{numerical framework} to address issues such as beam re-divergence in thick samples.\\\indent
The excellent agreement between the simulated and the experimental data is a critical first step towards real-space image reconstruction based on the current approach. State-of-the-art phase retrieval methods rely on Fourier transform to propagate wave back and forth between the sample and the detector space, which limits their application strictly in the kinematical limit. In the case of BCDI, non-negligible dynamical effects have been demonstrated for Au particles of as small as a few hundred nm in size \cite{Shabalin2017DynamicalCrystals, Hu2018DynamicImaging, Gao2022}. Despite showing how a TTE based approach could potentially account for those effects in a simulated scenario, the authors did not provide any solution to the inverse problem (\textit{i.e.}, from intensity to phase), due to the lack of an analytical form based upon which the wave can be back-propagated from the detector space to the sample space. Taking advantage of the resemblance of the numerical framework (Fig.~\ref{fig:sim_sketch}) to a multilayer perceptron, we circumvent this requirement of wave back-propagation by using automatic differentiation based gradient descent optimization, a method commonly used for the training of a NN. We demonstrate the essence of this capability (\textit{i.e.,} from intensity to strain) by retrieving both the lattice strain and the thickness of the LSMO film from a single diffraction pattern. Our results thus establish a foundation upon which reconstruction methods can be built to spatially resolve strain within 3D volumes of crystals from thousands of diffraction patterns measured at distinct overlapping sample locations in a manner analogous to Bragg ptychography. Such an implementation is particularly important for samples where substrate influence on the diffraction intensity is prominent.\\\indent

\begin{acknowledgments}
This work was performed at the Center for Nanoscale Materials. Use of the Center for Nanoscale Materials and Advanced Photon Source, both Office of Science user facilities, was supported by the U.S. Department of Energy, Office of Science, Office of Basic Energy Sciences, under Contract No. DE-AC02-06CH11357. Work related to synthesis of the material was supported by the National Science Foundation under grant No. DMR - 1745450. We thank Anatoly Shabalin and Hanfei Yan for the useful discussions.
\end{acknowledgments}

\review{
\section*{Appendices}
\subsection*{Appendix A: Order of Computation for the Numerical Framework}
\begin{figure}[ht]
\centering
\includegraphics[width=1\textwidth]{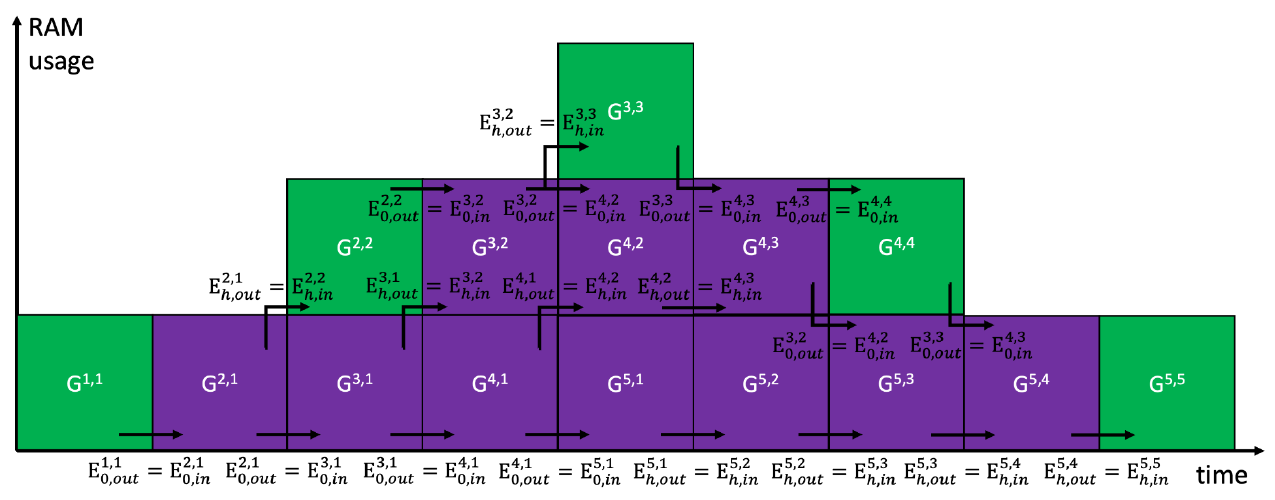}
\caption{\label{fig:order} \review{Order of computation versus RAM usage for $M=5$. The boundary conditions between neighbouring subsections are listed explicitly.}}
\end{figure}
Breaking down the integration network into smaller subsections allows the calculation of an exceedingly large sample volume with relatively low computer memory (RAM) usage. This is particularly useful when dynamical diffraction from the substrate needs to be considered, in which case the numerical integration is performed for \review{a minimum sample thickness that equals} the extinction depth of the substrate material ($\sim 10 ~\mu m$). Figure~\ref{fig:order} shows the order of computation in the case of $M=5$. The green blocks are subsections containing the sample surface while the purple blocks are subsections containing the bulk substrate. The computation is mostly sequential as limited by the boundary conditions, although a certain level of parallelism can be achieved.
}

\review{
\subsection*{Appendix B: Effect of coupling between the film and the substrate}
\begin{figure}[ht]
\centering
\includegraphics[width=1\textwidth]{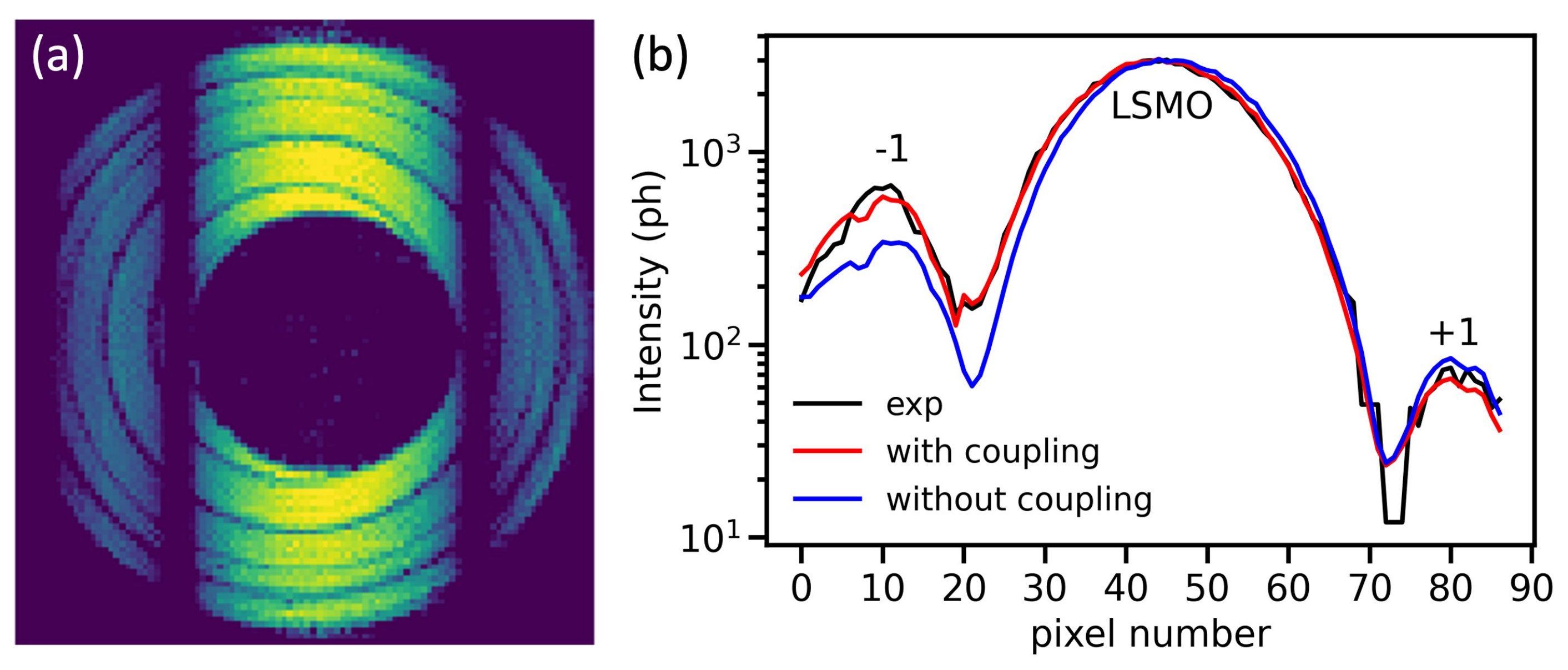}
\caption{\label{fig:coupling} \review{Effect of coupling between the film and the substrate. (a) Simulated far-field diffraction pattern at $\theta=\theta_{\rm{LSMO}}$ without contribution from the substrate. (b) Comparison between the experimental data (black), simulated intensity with (red curve) and without (blue curve) the coupling between the thin film and the substrate. The three peaks are respectively the -1 order thickness fringe, the 002 LSMO film Bragg peak and the +1 order thickness fringe. The dip in the experimental intensity at about pixel number 72 was due to missing intensities in the gap between the detector modules.}}
\end{figure}
With the approach proposed in this work, the coupling between the film and the substrate is considered by simply placing the film on top of the substrate in the numerical framework. The coherent sum of the intensity diffracted by the film and by the substrate is naturally included in the simulation, without the need to explicitly state the nature of their coupling. Fig.~\ref{fig:coupling}a was obtained by considering just the thin film in the numerical framework. The black curve in Fig.~\ref{fig:coupling}b was obtained by summing up the experimental detector intensity in Fig.~\ref{fig:film_peak}a in the vertical direction. The red curve was obtained by summing up the simulated detector intensity in Fig.~\ref{fig:film_peak}b in the vertical direction. To obtain the intensity profile without the substrate-film coupling, we first simulate separately the film diffraction pattern and the substrate diffraction pattern with consideration of dynamical effects. The intensities from the two diffraction patterns were added together incoherently. The blue curve was then obtained by summing up the resulted diffraction pattern in the vertical direction. An intensity shift of 1 pixel in Fig.~\ref{fig:coupling}b corresponds to a peak shift of 0.003$^\circ$ in 2$\theta$.}

\begin{figure}[ht]
\centering
\includegraphics[width=1\textwidth]{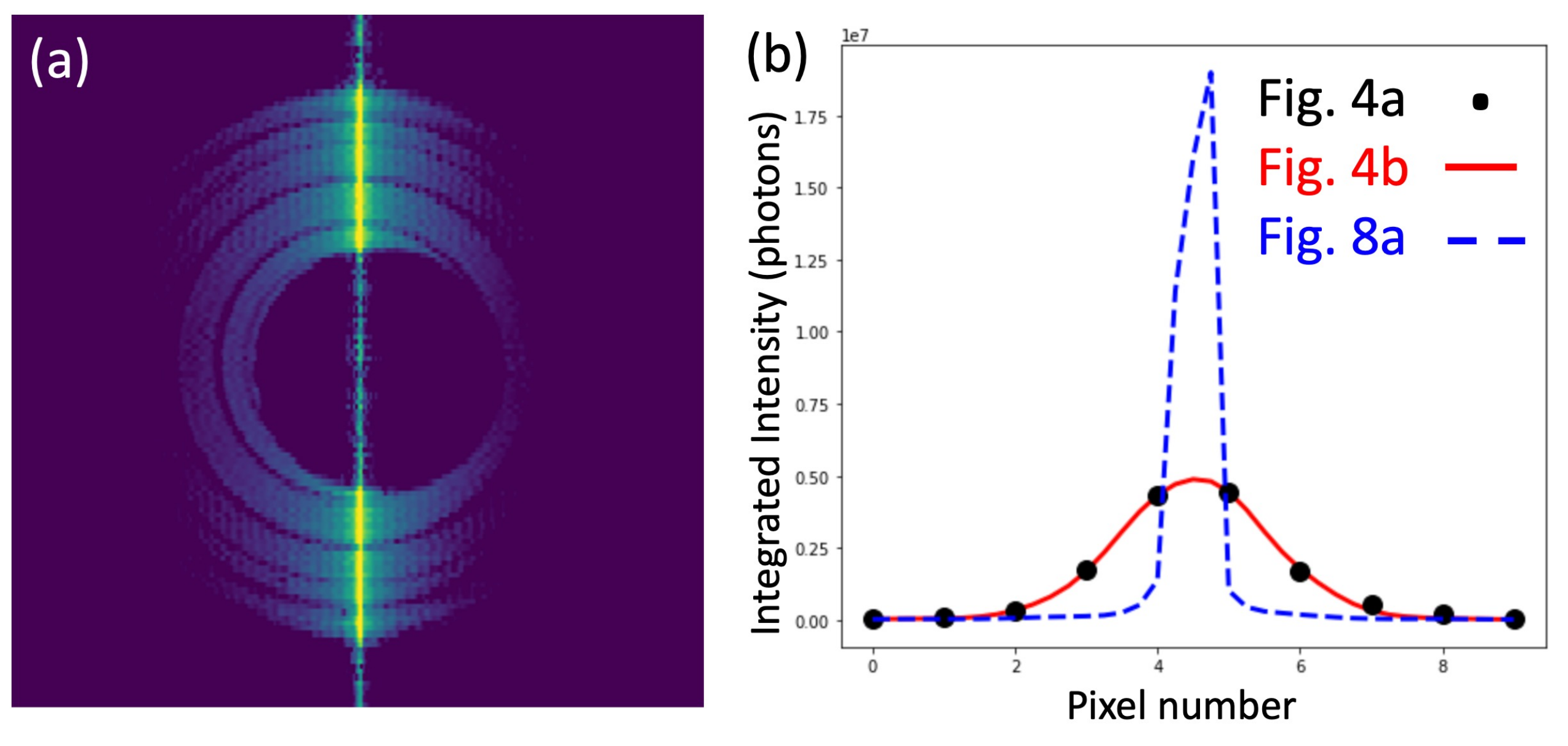}
\caption{\label{fig:broadening}  (a) Simulated far-field diffraction pattern at $\theta=\theta_{\rm{STO}}$. The breadth of the Bragg streak was given by dynamical diffraction effects. (b) Width of the substrate Bragg peak in number of pixels. Dynamical effects (blue) alone cannot explain the peak broadening observed in the experimental data (black).}
\end{figure}

\review{
\subsection*{Appendix C: Peak Broadening by Substrate Mosaicity}
Fig.~\ref{fig:broadening}a shows the diffraction pattern of a bare STO substrate without the consideration of substrate mosaicity. As shown in Fig.~\ref{fig:broadening}b, the peak breadth of the dynamically diffracting substrate peak (blue curve) is significantly narrower than the experimental data (black dots). To account for this, a total of 31 LSMO on STO simulations were performed, each differing relative to one another by a 0.001$^\circ$ rotation of the substrate in the scattering plane centered at $\theta_{\rm{STO}}$. The intensity of these 31 diffraction patterns were weighted by a Gaussian distribution function and summed incoherently. The FWHM of the Gaussian distribution function was 0.008$^\circ$, corresponding to the mosaicity spread of the STO crystal which was determined by lab x-ray diffraction on the same substrate prior to the thin film growth. The resulted peak breadth (red curve) was in excellent agreement with the experimental data.}

\end{document}